\documentclass[aps,prd,twocolumn,groupedaddress,amssymb,nofootinbib]{revtex4-1}
\usepackage{amsmath}
\usepackage{graphicx}
\usepackage{bbm}
\usepackage{xcolor,footnote} 
\begin{document}

\title{Baryon bags in strong coupling QCD}

\author{Christof Gattringer}
\email[]{christof.gattringer@uni-graz.at}

\affiliation{Universit\"{a}t Graz, Institut f\"{u}r Physik, Universit\"{a}tsplatz 5, 8010 Graz, Austria}

\date{26.2.2018}

\begin{abstract}
We discuss lattice QCD with one flavor of staggered fermions and show that in the path integral 
the baryon contributions can be fully separated from quark and diquark contributions. The baryonic degrees of freedom 
are independent of the gauge field and the corresponding free fermion action describes the baryons through the joint 
propagation of three quarks. The non-baryonic dynamics is described by quark and diquark terms that couple to the gauge field.
When evaluating the quark and diquark contributions in the strong coupling limit the partition function completely factorizes 
into baryon bags and a complementary domain. Baryon bags are regions in space-time where the dynamics is described 
by a single free fermion made out of three quarks propagating coherently as a baryon. Outside the baryon bags the 
relevant degrees of freedom are monomers and dimers for quarks and diquarks. The partition sum is a sum over all 
baryon bag configurations and for each bag a free fermion determinant appears as a weight factor. 
\end{abstract}

\maketitle

\section{Introduction}

The path integral of a quantum field theory usually can be represented in several ways that may highlight different properties
of the theory or allow for different computational approaches. In particular in the framework of the lattice discretization an optimal 
representation may considerably improve numerical simulations or even make the numerical approach applicable at all. Experience
from many years of Monte Carlo simulations of lattice field theories and spin systems shows that a representation that allows for an
efficient numerical simulation has to capture the relevant degrees of freedom such that these can be updated efficiently.

In recent years new worldline and worldsheet representations of lattice field theories were studied intensively, and the reviews at 
the annual lattice conferences give an overview over this development \cite{rev1,rev2,rev3,rev4}. For a QCD-related system 
a worldline representation
has been known since the early days of lattice QCD, namely the loop-dimer-monomer representation of strong coupling QCD 
\cite{RossiWolff,KarschMutter}. More recently several interesting results were obtained for strong coupling QCD 
\cite{scQCD1,scQCD2,scQCD3,scQCD4,scQCD5,scQCD6,scQCD7,scQCD8,scQCD9,scQCD10,scQCD11} 
with this type of representations and related expansion techniques for fermions were also used in various suggestions for a fully 
dualized version of lattice QCD \cite{GM2,Gagliardi,GM3,Borisenko,GM4}.

Worldline techniques also play a prominent role in purely fermionic lattice field theories with fermionic self-interaction terms. For some
of these theories it is possible to work with so-called fermion bags, which are domains on the lattice where the dynamics is essentially
described by free fermions. The fermion bags are separated by domains where the Grassmann integral for the partition function is saturated 
with the interaction terms of the theory \cite{FBrev1,FBrev2}. For systems where a fermion bag representation is available this constitutes a
powerful tool for numerical simulations and a wealth of interesting results were obtained with this approach 
\cite{FBres1,FBres2,FBres3,FBres4,FBres5,FBres6,FBres7,FBres8,FBres9,FBres10,FBres11}. 

In this article we show that also lattice QCD allows for a representation where the baryonic degrees of freedom are described by 
free fermions, i.e., lattice QCD can be organized in a generalized fermion bag formulation. Of course in QCD these are not fundamental 
fermions but the baryon contributions are terms where three quarks propagate together. One finds that the corresponding terms are 
independent of the gauge field and only the non-baryonic quark and diquark contributions interact with the gauge field. 

When considering the strong coupling limit the equivalence with the fermion bag picture becomes complete. We show that the path 
integral completely factorizes into regions ${\cal B}_i$, which we refer to as baryon bags, where the dynamics is described by free 
baryons made out of three quarks and a complementary domain $\overline{{\cal B}}$. In the complementary domain quarks and diquarks
are the relevant degrees of freedom. The form of the baryon bag representation which we find for the strong coupling limit can be viewed
as a variant of the loop-dimer-monomer representation \cite{RossiWolff,KarschMutter}, where now classes of 
contributions are resummed into the baryon bags. We show that the corresponding weights can be written as fermion determinants for a
free Dirac operator restricted to the respective baryon bag.

\section{Separation of the baryons in the QCD path integral}

We consider lattice QCD with one flavor of staggered fermions. The corresponding partition 
sum is given by 
\begin{eqnarray}
&& Z \;  = \; \int \!\! D [\, \overline{\psi}, \psi ] \int \!\! D[U]\, e^{\,S_G[U]} \, e^{\,S_F[\, \overline{\psi}, \psi, U]}  \; ,
\nonumber \\
&& \int \!\! D[U] \;  = \;  \prod_{x,\nu} \; \int_{\mbox{\tiny{SU(3)}}} \!\!\!\! d U_{x,\nu}  \; , 
\nonumber \\
&& \int \! D[\, \overline{\psi}, \psi] \; = \;  \prod_{x} \int \!\prod_{a=1}^3 d \psi_{x,a} \, d \overline{\psi}_{x,a}  \; .
\end{eqnarray}
The fermionic degrees of freedom are the one-component Grassmann valued fields 
$\psi_{x,a}$ and $\overline{\psi}_{x,a}$, with a color index $a = 1,2,3$. They live on the sites of a $N_s^3 \times N_t$ 
lattice and have
periodic boundary conditions in space and anti-periodic boundary conditions in time. The corresponding path integral
measure is a product over Grassmann measures for all sites and colors. 
The gauge degrees of freedom are the matrices $U_{x,\nu} \in $ SU(3), attached to the links of the 
lattice. They obey periodic boundary conditions for all directions. The corresponding path integral measure is the product 
over Haar measures on all links.

For the gauge action we use the Wilson action (the constant term was dropped)
\begin{equation}
S_G[U] \; = \; \frac{\beta}{3} \sum_x \sum_{\mu < \nu} \mbox{Re} \, 
\mbox{Tr} \, U_{x,\mu} \, U_{x+\hat{\mu},\nu} \, U_{x+\hat{\nu},\mu}^\dagger \, U_{x,\nu}^\dagger \; .
\end{equation}
The staggered fermion action is given by
\begin{eqnarray}
&& \hspace{15mm} S_{F}\big[\overline{\psi}, \psi, U\big]  =  \sum_{x} \biggl( 2m \, \overline{\psi}_x \psi_x \; +
\label{quarkaction}
\\
&&\sum_{\nu} \gamma_{x,\nu} \Big[ \, e^{\mu \delta_{\nu 4}} \, \overline{\psi}_x U_{x,\nu} \psi_{x + \hat{\nu}}  - 
e^{-\mu \delta_{\nu 4}} \, \overline{\psi}_{x + \hat{\nu}} U_{x,\nu}^{\dagger} \psi_x \Big] \!\biggr) \; ,
\nonumber
\end{eqnarray}
where we use vector/matrix notation in color space. The staggered sign factors are given by
$\gamma_{x,1} = 1$, $\gamma_{x,2} = (-1)^{x_1}$, $\gamma_{x,3} = (-1)^{x_1+x_2}$ and $\gamma_{x,4} = (-1)^{x_1+x_2+x_3}$.
For later discussion we have also included a chemical potential $\mu$ which gives a different weight to
temporal ($\nu = 4$) forward and backward hopping of the fermions. 
 
Using the fact that the fermion action consists only of Grassmann bilinears, we may write the sums in the 
exponent as products and organize the partition function in the form
\begin{equation}
Z = \int \!\! D [\, \overline{\psi}, \psi ] \prod_x \!e^{2m \, \overline{\psi}_x \psi_x } \!\!\! \int \!\! D[U]\, e^{\,S_G[U]}
\prod_{x,\nu}  L_{x,\nu}[ \, \overline{\psi}, \psi,U ] ,
\end{equation}
where we have defined the link terms
\begin{eqnarray}
L_{x,\nu}[ \, \overline{\psi}, \psi,U ]  & = &   
e^{\, \gamma_{x,\nu} \, e^{\mu \delta_{\nu 4}} \, \overline{\psi}_x U_{x,\nu} \psi_{x + \hat{\nu}} } 
\\
&& \hspace{11mm} \times \;
e^{\, -\gamma_{x,\nu} \, e^{-\mu \delta_{\nu 4}} \, \overline{\psi}_{x + \hat{\nu}} U_{x,\nu}^{\dagger} \psi_x } \; .
\nonumber 
\end{eqnarray}
Expanding the exponential for the forward hopping term in a power series we find (for brevity we 
drop all space-time indices; note that $\gamma \equiv \gamma_{x,\nu}$ is a sign such that it appears only in odd powers) 
\begin{eqnarray}
&& e^{\, \gamma \, e^{\mu} \, \overline{\psi} \, U \psi } = 
1  + \gamma \, e^{\mu} \; \overline{\psi} U \psi \,  + \frac{e^{2\mu}\!}{2 !} (\overline{\psi} U \psi)^2 
 + \frac{\gamma \,e^{3\mu} \!\!}{3 !} (\overline{\psi} U \psi)^3 
\nonumber \\
& & \;\; = \;
\Big[ 1 + \frac{\gamma \,e^{3\mu}}{3 !} (\overline{\psi} U \psi)^3 \Big] 
\Big[1  + \gamma \, e^{\mu} \; \overline{\psi} U \psi \, + \frac{e^{2\mu}}{2 !} (\overline{\psi} U \psi)^2 \Big]
\nonumber \\
& & \hspace{7mm} =  \; \exp\left( \frac{\gamma \,e^{3\mu}}{3 !} (\overline{\psi} U \psi)^3 \right) \, \sum_{k=0}^2 
\frac{ \Big[ \gamma \, e^{\mu} \; \overline{\psi} U \psi \Big]^k}{k!} 
\; .
\label{aux1}
\end{eqnarray}
Due to the nilpotency of the $\psi = \psi_x$ and the $\overline{\psi} = \overline{\psi}_x$ and the fact that there are 
three colors, the power series terminates after the cubic term. 
Again exploiting nilpotency we have separated the contribution of the cubic term in the second step and in the 
last step re-exponentiated that factor. The cubic term can be written as (we use summation convention for the color 
indices)
\begin{eqnarray}
(\overline{\psi} U \psi)^3  & \, = \, & (\overline{\psi}_a U_{ab} \psi_b)^3 
\label{aux1}
\\
&  \, = \, & 
\overline{\psi}_a \, \psi_b \, \overline{\psi}_{a^\prime} \, \psi_{b^\prime} \, 
\overline{\psi}_{a^{\prime\prime}} \, \psi_{b^{\prime\prime}} \;
U_{ab} \, U_{a^\prime b^\prime} \, U_{a^{\prime\prime} b^{\prime\prime}} 
\nonumber \\
& \, = \, & 
- \, \overline{\psi}_a \, \overline{\psi}_{a^\prime} \, \overline{\psi}_{a^{\prime\prime}} \, 
\psi_b \,  \psi_{b^\prime} \, \psi_{b^{\prime\prime}} \,
U_{ab} \, U_{a^\prime b^\prime} \, U_{a^{\prime\prime} b^{\prime\prime}} 
\nonumber \\
& \, = \, &
\overline{\psi}_3 \, \overline{\psi}_{2} \, \overline{\psi}_{1} \, \psi_1 \,  \psi_{2} \, \psi_{3} \, 
\epsilon_{a \, a^\prime a^{\prime\prime}}\, 
U_{ab} \, U_{a^\prime b^\prime} \, U_{a^{\prime\prime} b^{\prime\prime}} \,
\epsilon_{b \, b^\prime b^{\prime\prime}} 
\nonumber \\
& \, = \, & \overline{\psi}_3 \, \overline{\psi}_{2} \, \overline{\psi}_{1} \, \psi_1 \,  \psi_{2} \, \psi_{3} \;
\epsilon_{a \, a^\prime a^{\prime\prime}}\, \epsilon_{a \, a^\prime a^{\prime\prime}}\, \det U 
\nonumber \\
& \, = \, & 
3! \; \overline{\psi}_3 \, \overline{\psi}_{2} \, \overline{\psi}_{1} \, \psi_1 \,  \psi_{2} \, \psi_{3} \; \equiv  \; 
3! \; \overline{B} \, B \; .
\nonumber
\end{eqnarray} 
In the last two steps we used $U_{ab} \, U_{a^\prime b^\prime} \, U_{a^{\prime\prime} b^{\prime\prime}} \,
\epsilon_{b \, b^\prime b^{\prime\prime}} =  \epsilon_{a \, a^\prime a^{\prime\prime}} \det U$ and 
$\det U = 1$, since $U \in$ SU(3). The calculation shows that the cubic term is independent of the gauge field, and 
for a compact notation in the last line of (\ref{aux1}) we already used the baryon fields $B_x$, 
$\overline{B}_x$ defined as
\begin{equation}
B_x \; = \;  \psi_{x,1} \, \psi_{x,2} \, \psi_{x,3} \;, \qquad 
\overline{B}_x \; = \;  \overline{\psi}_{x,3} \, \overline{\psi}_{x,2} \, \overline{\psi}_{x,1} \; .
\label{baryondefs}
\end{equation}

Performing the equivalent steps for the backward hopping term and reinstating all indices we can write the 
link terms in the form
\begin{eqnarray}
&& L_{x,\nu}[ \, \overline{\psi}, \psi,U]  \; = \; e^{ \gamma_{x,\nu} [ e^{\, 3\mu \delta_{\nu 4}} \, 
\overline{B}_x  B_{x + \hat{\nu}}  - 
e^{-3\mu \delta_{\nu 4}} \, \overline{B}_{x + \hat{\nu}} B_x ]}  
\nonumber \\
&&
\times \sum_{k_{x,\nu},j_{x,\nu} =0}^2 
\frac{(\gamma_{x,\nu})^{k_{x,\nu}+j_{x,\nu}} \, e^{ \, \mu \, \delta_{\nu 4}  \, (k_{x,\nu}-j_{x,\nu}) } }{k_{x,\nu}! \; j_{x,\nu}!}
\nonumber \\
&&
\times \;  
\Big[  \overline{\psi}_x \, U_{x,\nu} \, \psi_{x + \hat{\nu}} \Big]^{k_{x,\nu}} \;
\Big[ - \overline{\psi}_{x  +  \hat{\nu}} \, U_{x,\nu}^\dagger \, \psi_x \Big]^{j_{x,\nu}} \; .
\label{onelinkaux}
\end{eqnarray}
Obviously we have completely separated the contributions of the baryons to the link terms. 
These contributions have the form of a Boltzmann factor with the massless staggered fermion action for the baryon fields $\overline{B}_{x}$
and $B_x$. Only the terms in the double sum still depend on the gauge fields. They have the form of hopping terms
for quarks ($k_{x,\nu},j _{x,\nu}=1$) or hopping terms for diquarks ($k_{x,\nu},j_{x,\nu}=2$).

The Boltzmann factors for the mass term can be organized in a similar way
(again we suppress space-time indices and 
use $(\overline{\psi} \psi)^3 = (\overline{\psi}_1 \psi_1 +
\overline{\psi}_2 \psi_2 + \overline{\psi}_3 \psi_3)^3 =  3! \,
\overline{\psi}_3 \overline{\psi}_2 \overline{\psi}_1 \psi_1 \psi_2 \psi_3  = 3! \, 
\overline{B} \, B\,$):
\begin{eqnarray}
e^{2m \, \overline{\psi} \psi} & = &  
1 \; + \; 2m \,  \overline{\psi} \psi \; + \; \frac{(2m)^2}{2 !} (\overline{\psi} \psi)^2 \; + \; 
(2m)^3 \,  \overline{B} \, B 
\nonumber \\
& = & 
\Big[1 \, + \, (2m)^3 \,  \overline{B} \, B  \Big]
\Big[ 1 \, + \, 2m \,  \overline{\psi} \psi \, + \, 
\frac{(2m)^2}{2 !} (\overline{\psi} \psi)^2 \Big] \nonumber \\
& = & e^{\, (2m)^3 \,  \overline{B} \, B } 
\Big[ 1 \, + \, 2m \,  \overline{\psi} \psi \, + \, \frac{(2m)^2}{2 !} (\overline{\psi} \psi)^2 \Big] \; 
 \; .
\end{eqnarray}
Reinstating all space time-factors and using the definitions (\ref{baryondefs}) we find
\begin{equation}
e^{\, 2m \, \overline{\psi}_x \psi_x} \; = \; e^{\, 2 M \; \overline{B}_x B_x} \,
\sum_{s_x = 0}^2 \frac{ \Big[ 2m \,  \overline{\psi}_x \psi_x \Big]^{s_x}}{s_x !} \; ,
\label{massterms}
\end{equation}
where we defined the baryon mass $M = 4 m^3$, and introduced the monomer variables $s_x = 0,1,2$ assigned to the sites $x$ of the lattice.

Putting things together we obtain the following expression for the partition sum:
\begin{equation}
Z  = \int \!\! D [\, \overline{\psi}, \psi ] \, e^{ \, S_B[\, \overline{B},B]} \!  \int \!\! D[U]\, e^{\,S_G[U]} \, 
W_{QD} [\, \overline{\psi}, \psi, U] \; .
\label{Zfactor}
\end{equation}
In this form we have completely factorized the contributions of the baryons and collected the corresponding terms in the baryon action
\begin{eqnarray}
&& \hspace{0mm} S_{B}\big[ \, \overline{B}, B \big]  =  \sum_{x} \biggl( 2M \, \overline{B}_x B_x \; +
\label{baryonaction}
\\
&&\sum_{\nu} \gamma_{x,\nu} \Big[ \, e^{\mu_B \delta_{\nu 4}} \, \overline{B}_x \, B_{x + \hat{\nu}}  - 
e^{-\mu_B \delta_{\nu 4}} \, \overline{B}_{x + \hat{\nu}}  B_x \Big] \!\biggr) \; .
\nonumber
\end{eqnarray}
Obviously the action for the baryon fields $B_x$ and $\overline{B}_x$ 
has the form of a free staggered fermion with mass $M = 4 m^3$ and a baryon chemical potential $\mu_B = 3 \mu$. 
Note, that since the anti-periodic temporal boundary conditions of the quark fields simply correspond to 
additional factors of $-1$ for the temporal links of the last time slice in the action (\ref{quarkaction}), 
also the baryon fields $B_x$ and $\overline{B}_x$ obey anti-periodic temporal boundary conditions 
(the corresponding signs $\pm 1$ are raised to a power of 3 in our derivation and because of 
$(\pm 1)^3 = \pm 1$, remain unchanged). Furthermore we point out that also the baryon action 
turns into a chiral action (chirality in the sense of the staggered action) for the limit of massless quarks.

The last factor in (\ref{Zfactor}), i.e., the expression $W_{QD} [\, \overline{\psi}, \psi, U]$ is the integrand with the 
contributions of the non-baryonic terms. It reads
\begin{eqnarray}
&& \hspace{-5mm} W_{QD} [\, \overline{\psi}, \psi, U] \; = \; \sum_{\{s,k,l\}}  
\prod_{x} \frac{ \Big[ 2m \,  \overline{\psi}_x \psi_x \Big]^{s_x}\!\!}{s_x !} \; 
\label{WQD} \\
&& \qquad
\times \; \prod_{x,\nu} 
\frac{(\gamma_{x,\nu})^{k_{x,\nu}+l_{x,\nu}} \, e^{ \, \mu \, \delta_{\nu 4}  \, (k_{x,\nu}-l_{x,\nu}) } }{k_{x,\nu}! \; l_{x,\nu}!}
\nonumber \\
&& \qquad
\times \;  
\Big[  \overline{\psi}_x \, U_{x,\nu} \, \psi_{x + \hat{\nu}} \Big]^{k_{x,\nu}} \;
\Big[ - \overline{\psi}_{x  +  \hat{\nu}} \, U_{x,\nu}^\dagger \, \psi_x \Big]^{l_{x,\nu}} \; ,
\nonumber
\end{eqnarray}
where we have defined the sum over all configurations of the monomer and hopping variables as
\begin{equation}
\sum_{\{s,k,l\}}  \; = \; \left[ \prod_x \sum_{s_x = \,0}^2 \right]
\left[ \prod_{x,\nu} \, \sum_{k_{x,\nu} = \,0}^2 \; \sum_{l_{x,\nu} = \,0}^2 \right] \; .
\end{equation}
The contributions to $W_{QD} [\, \overline{\psi}, \psi, U]$ have a simple interpretation as terms for quarks and diquarks.
More specifically, a value $s_x = 1$ inserts a quark monomer at site $x$, while $s_x = 2$ inserts a diquark monomer at site $x$.
In the same way a value of $k_{x,\nu} = 1$ activates a forward quark hop on the link $(x,\nu)$, while a 
value of $k_{x,\nu} = 2$ activates a forward hop for a diquark. Finally, setting $l_{x,\nu} = 1$ and $l_{x,\nu} = 2$ activate backward hops 
for quarks and diquarks. 

Thus we conclude that in the form (\ref{Zfactor}) the partition sum factorizes into the contributions of gauge independent, freely propagating 
baryons described by the baryon action $S_B[\, \overline{B},B]$ and the contributions $W_{QD} [\, \overline{\psi}, \psi, U]$ 
of quarks and diquarks that couple to the gauge fields. We expect that the representation (\ref{Zfactor}) will be useful for simplifying dual 
representations of lattice QCD in terms of worldlines or worldsheets -- a task we will address in future work.

\section{Integrating the gauge fields at strong coupling}

We now evaluate (\ref{Zfactor}) in the strong coupling limit. This means that the gauge action is absent and we integrate the gauge variables 
$U_{x,\nu}$ at every link without weight factor. Thus we can completely factorize the integration over the link variables and  
for strong coupling the partition function reads
\begin{eqnarray}
&& Z =  \int \!\! D [\, \overline{\psi}, \psi ] \, e^{ \, S_B[\, \overline{B},B]} \;
\prod_{x} \sum_{s_x = 0}^2 \frac{ \Big[ 2m \,  \overline{\psi}_x \psi_x \Big]^{s_x}\!\!}{s_x !} \;
\label{aux3}
\\
&& \qquad
\times \prod_{x,\nu} \sum_{k_{x,\nu} = 0}^2 \sum_{l_{x,\nu} = 0}^2
\frac{(\gamma_{x,\nu})^{k_{x,\nu}+l_{x,\nu}} \, e^{ \, \mu \, \delta_{\nu 4}  \, (k_{x,\nu}-l_{x,\nu}) } }{k_{x,\nu}! \; l_{x,\nu}!}
\nonumber \\
&& \qquad 
\times \! \int_{\mbox{\tiny{SU(3)}}} \!\!\!\!\!\! d U_{x,\nu}
\Big[  \overline{\psi}_x \, U_{x,\nu} \, \psi_{x + \hat{\nu}} \Big]^{k_{x,\nu}} \;
\Big[ - \overline{\psi}_{x  +  \hat{\nu}} \, U_{x,\nu}^\dagger \, \psi_x \Big]^{l_{x,\nu}} .
\nonumber
\end{eqnarray}
Integrating the $U_{x,\nu}$ in the last line of (\ref{aux3}) over only 
the center group already restricts the possible values of $k_{x,\nu}$ and $l_{x,\nu}$ to $k_{x,\nu}=l_{x,\nu}$. 
This leaves us with two simple well known SU(3) Haar measure integrals \cite{creutz}:
\begin{eqnarray}
&& \int_{\mbox{\tiny{SU(3)}}} \!\!\!\!\!\!\!\!\!\! d U \, U_{ab} \, U^\dagger_{cd} = \frac{1}{3} \, \delta_{ad} \, \delta_{bc} \, ,
 \\
&& \int_{\mbox{\tiny{SU(3)}}} \!\!\!\! \!\!\!\!\!\! d U \, U_{ab} \, U_{cd} \, U^\dagger_{ef} \, U^\dagger_{gh} =  
\frac{1}{8} [ \delta_{af} \, \delta_{be} \, \delta_{ch} \, \delta_{dg}  +  \delta_{ah} \, \delta_{bg} \, \delta_{cf} \, \delta_{de} ]
\nonumber \\
&& \hspace{31mm} 
- \; \frac{1}{24} [ \delta_{af} \, \delta_{bg} \, \delta_{ch} \, \delta_{de} +  \delta_{ah} \, \delta_{be} \, \delta_{cf} \, \delta_{dg} ] .
\nonumber
\end{eqnarray}
With these formulas we obtain for the two non-trivial integrals in (\ref{aux3}),
\begin{eqnarray}
\int_{\mbox{\tiny{SU(3)}}} \!\!\!\!\!\!\!\!\!\! d U  \big[\,\overline{\psi}_x \, U \, \psi_{x + \hat{\nu}}\big] \!
\big[\,\overline{\psi}_{x  +  \hat{\nu}} U^\dagger \psi_x \big]  =  - \frac{\overline{\psi}_x  \psi_x \; \, 
\overline{\psi}_{x  +  \hat{\nu}} \psi_{x + \hat{\nu}}}{3},
\nonumber 
\\
\int_{\mbox{\tiny{SU(3)}}} \!\!\!\!\!\!\!\!\!\! d U  \big[\,\overline{\psi}_x \, U \, \psi_{x + \hat{\nu}}\big]^{\!2} 
\big[\,\overline{\psi}_{x  +  \hat{\nu}} U^\dagger \psi_x \big]^{\!2} \! =  \frac{ \big[\,\overline{\psi}_x  \psi_x \;\;
\overline{\psi}_{x  +  \hat{\nu}} \psi_{x + \hat{\nu}}\big]^{\!2}\!\!\!}{3},
\nonumber
\\
\label{twointegrals}
\end{eqnarray}
which agrees with the two leading non-trivial terms for the one link integral \cite{RossiWolff,KarschMutter}.  
Using (\ref{twointegrals}), the strong coupling partition function assumes the form
\begin{eqnarray}
&& \hspace{-4mm}
Z =  \int \!\! D [\, \overline{\psi}, \psi ] \, e^{ \, S_B[\, \overline{B},B]} \;
\prod_{x} \sum_{s_x = 0}^2 \frac{ \Big[ 2m \,  \overline{\psi}_x \psi_x \Big]^{s_x}\!\!}{s_x !} \;
\nonumber
\\
&&
\hspace{3mm}
\times \prod_{x,\nu}  \sum_{k_{x,\nu} =0}^2 \!\!\!  \frac{(3-k_{x,\nu})!}{6 \,  k_{x,\nu}!} \big[\,\overline{\psi}_x  \psi_x \;\;
\overline{\psi}_{x  +  \hat{\nu}} \psi_{x + \hat{\nu}}\big]^{k_{x,\nu}} \! .
\label{zstrong1}
\end{eqnarray}
The terms in the last sum correspond to quark dimers $(k_{x,\nu} = 1)$, i.e., a term where a quark hops one link forward and then backwards
on the same link, and to diquark dimers $(k_{x,\nu} = 2)$ where two quarks hop forward together and then backwards.

\section{Factorization of  the Grassmann integral and definition of baryon bags}

Having brought the strong coupling partition function into the form (\ref{zstrong1})
we now come to identfying the fermion bags. The important step is to notice,  
that already with only the Boltzmann factor $e^{\, S_B[\, \overline{B},B]}$ for the baryons we can 
saturate the Grassmann integral. Saturating the Grassmann integral means that when expanding 
the integrand we collect those terms of the expansion where each Grassmann variable
appears exactly once, such that the Grassmann integral gives a non-vanishing result.  

The action $S_B[\, \overline{B},B]$ in (\ref{baryonaction}) has the form of the staggered fermion action, 
and upon expansion the Boltzmann factor $e^{\, S_B[\, \overline{B},B]}$ produces the same terms: 
monomers, dimers and closed loops, which we can use to saturate the Grassmann integral. Note
however, that the monomer, dimer and loop contributions are now for the baryons, and thus already contain 
all three colors, since the baryons  $B_x$ and $\overline{B}_x$ defined in (\ref{baryondefs}) are products 
over all three colors. Thus if we have a region ${\cal B}_i$ of the lattice where every site is either occupied 
by a baryon monomer, is the endpoint of a baryon dimer, or is run through by a baryon loop, then we 
have complete saturation of the Grassmann integral inside the region ${\cal B}_i$. 

\begin{figure}[t]
\begin{center}
\includegraphics[height=5cm]{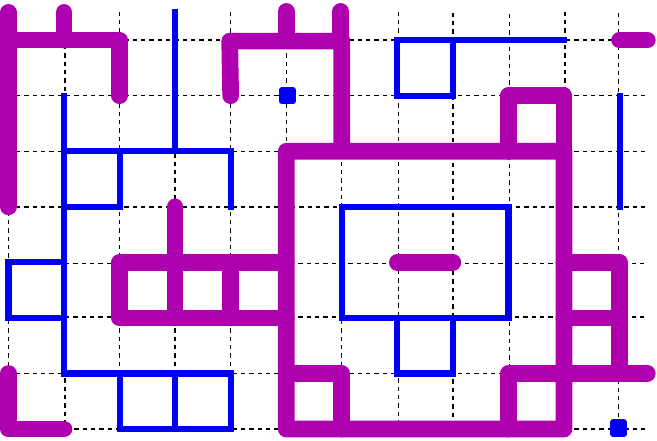}
\end{center}
\caption{Example for the decomposition of a lattice into baryon bags ${\cal B}_i$ and the complementary region
$\overline{ {\cal B}}$. The baryon bags are marked with the thinner blue lines on the links they contain or by a 
blue square if they consist of only a single site. The links in the complementary region $\overline{ {\cal B}}$ are marked with 
thick magenta lines. The configuration we show has 6 baryon bags ${\cal B}_i$, 2 of which consist of only a 
single site. Inside each baryon bag the Grassmann integral is saturated by baryon monomers, baryon dimers and 
baryon loops. In the complementary region $\overline{ {\cal B}}$ the Grassmann integral is 
saturated by monomers and dimers for quarks and diquarks.}
\label{bag_example}
\end{figure}

Such a region ${\cal B}_i$ is now referred to as baryon bag. More specifically we define baryon bags 
${\cal B}_i$ to consist of either a single site (then the only possibility to saturate the Grassmann integral 
is the baryon monomer), or of the sites at the endpoints of a connected set of links. Examples of Baryon 
bags are shown in Fig.~\ref{bag_example}. Inside the space time domain defined by a baryon bag we completely 
saturate the Grassmann integral with the contributions from $e^{\, S_B[\, \overline{B},B]}$, i.e., baryon monomers, 
baryon loops and baryon dimers.

Having introduced the fermion bags ${\cal B}_i$ as either a single site or a set of sites connected by links,
we define the union ${\cal B}$ of all bags and the complementary domain $\overline{{\cal B}}$,
\begin{equation}
{\cal B} \; \equiv \; \cup_{i} \, {\cal B}_i \; \; , \qquad \overline{{\cal B}} \; \equiv \; \Lambda \, / \, {\cal B} \; ,
\end{equation}
where $\Lambda$ denotes the full lattice, i.e., the set of all sites. On the complementary domain $\overline{{\cal B}}$ we can use 
quark monomers ($s_x = 1$), diquark monomers ($s_x = 2$), as well as quark dimers ($k_{x,\nu} = 1$), 
or diquark dimers ($k_{x,\nu} = 2$). We stress at this point that quark monomers ($s_x = 1$) and diquark monomers  ($s_x = 2$) 
are mutually exclusive, i.e., at a site one cannot have a quark monomer and a diquark monomer at the same instant, 
since they correspond to different values of the variable $s_x$ assigned to the site $x$. In the same way
quark dimers ($k_{x,\nu} = 1$) and diquark dimers ($k_{x,\nu} = 2$) are mutually excluding each other at a link $(x,\nu)$, 
since they correspond to different values of the variable $k_{x,\nu}$.

\begin{figure}[t]
\begin{center}
\includegraphics[height=5cm]{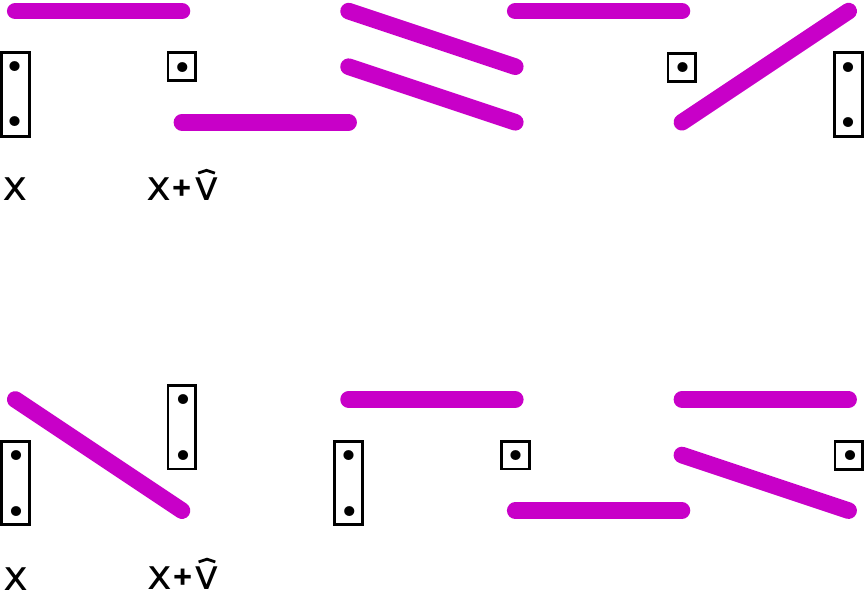}
\end{center}
\caption{Example for the saturation of the Grassmann integral in the complementary domain $\overline{{\cal B}}$. We show two
(top and bottom plots) of strings of neighboring sites, with the first site labelled by $x$, the second one by $x+\hat{\nu}$. At each site we 
have three layers for the three colors. The quark monomers  $\overline{\psi}_{x,a} \, \psi_{x,a}$ are represented by a small square 
around the site $x$ in the layer for color $a$ and the diquark monomers  $\overline{\psi}_{x,a} \, \psi_{x,a} \; \overline{\psi}_{x,b} \, \psi_{x,b}$
$(a \neq b)$ by a rectangle around $x$ in both layers $a$ and $b$. A quark dimer 
$\overline{\psi}_{x,a} \, \psi_{x,a} \; \overline{\psi}_{x+\hat{\nu},c} \, \psi_{x+\hat{\nu},c}$ is represented by a single magenta line
connecting $x$ in layer $a$ with $x+\hat{\nu}$ in layer $c$, while diquark dimers  $\overline{\psi}_{x,a} \, \psi_{x,a} \; \overline{\psi}_{x,b} \, \psi_{x,b} \;  
\overline{\psi}_{x+\hat{\nu},c} \, \psi_{x+\hat{\nu},c} \; \overline{\psi}_{x+\hat{\nu},d} \, \psi_{x+\hat{\nu},d}$ are represented
by two magenta lines that connect $x$ and $x+\hat{\nu}$ in the corresponding layers. For graphical reasons the two examples shown have the structure 
of chains, but of course the patterns for the saturation of the Grassmann integral can branch in all four directions.}
\label{complementary}
\end{figure}

The quark monomer terms generate the quadratic Grassmann terms $\overline{\psi}_{x,a} \, \psi_{x,a}, \, a = 1,2,3$, 
while the diquark monomer terms give rise to 
$\overline{\psi}_{x,a} \, \psi_{x,a}\; \overline{\psi}_{x,b} \, \psi_{x,b}$, $a \neq b$. Obviously neither quark-, nor diquark 
monomer terms alone can saturate the Grassmann integral at $x$, because they do not contain all three colors. Consequently,
the complementary region  $\overline{{\cal B}}$ cannot contain isolated sites. Thus, for configurations that saturate the 
Grassmann integral, the quark and diquark monomer terms have to be combined with quark and diquark hopping terms. These 
correspond to activating at the endpoints of the link $(x,\nu)$ the terms
$\overline{\psi}_{x,a} \, \psi_{x,a} \; \overline{\psi}_{x+\hat{\nu},c} \, \psi_{x+\hat{\nu},c}$ for the quark dimers, and
$\overline{\psi}_{x,a} \, \psi_{x,a} \; \overline{\psi}_{x,b} \, \psi_{x,b} \;  
\overline{\psi}_{x+\hat{\nu},c} \, \psi_{x+\hat{\nu},c} \; \overline{\psi}_{x+\hat{\nu},d} \, \psi_{x+\hat{\nu},d}$ for the diquark dimers.
Thus the configurations that saturate the Grassmann integral in the complementary domain $\overline{{\cal B}}$ consist of regions that combine 
monomers and dimers for quarks and diquarks. Some examples of such configurations are shown in Fig.~\ref{complementary}.

Having discussed the ways baryon terms from the expansion of $e^{\, S_B[\, \overline{B},B]}$ may be used to saturate the 
Grassmann integral inside the baroyn bags ${\cal B}_i$, and how quark and diquark monomers and dimers saturate the Grassmann 
integral in the complementary domain $\overline{{\cal B}}$, let us stress that the two types of contributions cannot mix at sites $x$: 
Clearly a baryon mass term completely saturates the Grassmann integral at some site $x$ and quark or diquark terms cannot attach 
to that site $x$. The second building block for baryons are the hopping terms for baryons. At the endpoint $x$ of a link they
either completely saturate all $\psi_{x,a}$ or all $\overline{\psi}_{x,a}$. The quark and diquark monomer and dimer terms on the other hand
always contain the products $\overline{\psi}_{x,a} \, \psi_{x,a}$, such that they cannot attach to the site $x$. Consequently the baryon 
contributions do not mix with the quark and diquark terms.

The fact that baryon terms do not mix with quark and diquark terms allows us to completely factorize the strong coupling path integral. The
Grassmann measure is decomposed in the following way:
\begin{equation}
\int \!\!  D[\, \overline{\psi}, \psi ]  \; = \; 
\prod_i \int \!\!  D_{{\cal B}_i}[\, \overline{\psi}, \psi ] \; \times \; \int \!\!  D_{\,\overline{{\cal B}}\,}[\, \overline{\psi}, \psi ] \; ,
\end{equation}
where we have defined the measures in the fermion bags and the complementary domain as
\begin{eqnarray}
&& \int \!\!  D_{{\cal B}_i}[\, \overline{\psi}, \psi ]  \; = \; 
\prod_{x \in {\cal B}_i} \int \! \prod_a  d\psi_{x,a} \, d\overline{\psi}_{x,a} \; ,
\label{measure_bag}
\\
&&
\int \!\!  D_{\,\overline{{\cal B}}\,}[\, \overline{\psi}, \psi ] \; = \;  \prod_{x \in \overline{{\cal B}}} \;
\int \! \prod_a  d\psi_{x,a} \, d\overline{\psi}_{x,a} \; .
\label{measure_complement}
\end{eqnarray}
Since we completely saturate the Grassmann integral inside the fermion bags with the baryon terms, the sum over quark and diquark 
activation indices is restricted to the complementary domain $\overline{{\cal B}}$ and we define
\begin{equation}
\sum_{\{s,k \, \| \, \overline{{\cal B}}\}} \; = \; \prod_{x \in \overline{{\cal B}}} \; \sum_{s_x = 0}^2 \; \; \;
\prod_{(x,\nu) \in \overline{{\cal B}}} \; \sum_{k_{x,\nu} = 0}^2  \; .
\end{equation}
The partition function can now be written in the form
\begin{eqnarray}
Z  \; = \; \sum_{\{{\cal B}\}} \left[ \prod_i \; Z_{ {\cal B}_i} \right] \times \; Z_{\,\overline{{\cal B}}} \; ,
\end{eqnarray}
where the sum runs over all possible ways to decompose the lattice into a collection of fermion bags ${\cal B}_i$. 
The contribution of a baryon bag ${\cal B}_i$ is given by
\begin{eqnarray}
Z_{{\cal B}_i}  \; = \; \int \!\!  D_{{\cal B}_i}[\, \overline{\psi}, \psi ] \; \exp\left( \sum_{x,y} \overline{B}_x \, D^{(i)}_{\;x,y} \, B_y \right) \; ,
\end{eqnarray}
where we have introduced $D^{(i)}_{\;x,y}$ as the staggered Dirac operator for the baryons in the baryon bag ${\cal B}_i$,
\begin{eqnarray}
D^{(i)}_{\;x,y}  & = & \theta^{(i)}_x \; 2M \, \delta_{x,y}  
\label{bagoperator}
\\
& + & \sum_{\nu} \theta^{(i)}_{x,\nu} \gamma_{x,\nu} 
\Big[ \, e^{\mu_B \delta_{\nu 4}} \, \delta_{x+\nu,y} \; - \;  e^{-\mu_B \delta_{\nu 4}} \, \delta_{x,y+\nu} \Big] \; ,
\nonumber 
\end{eqnarray}
with the site and link support functions on the  baryon bag ${\cal B}_i$ defined as
\begin{eqnarray}
\theta^{(i)}_x & = & \left\{ \begin{array}{cl}
1 & \mbox{if} \; x \in {\cal B}_i \\
0 & \mbox{if} \; x \notin {\cal B}_i 
\end{array}
\right. \; ,
 \\
\theta^{(i)}_{x,\nu} & = & \left\{ \begin{array}{cl}
1 & \mbox{if} \; (x,\nu) \in {\cal B}_i \\
0 & \mbox{if} \; (x,\nu) \notin {\cal B}_i 
\end{array}
\right. \; .
\end{eqnarray}
Finally, the contribution of the complementary domain $\overline{{\cal B}}$ is given by
\begin{eqnarray}
&& Z_{\,\overline{{\cal B}}} \; = \!\!\!\! \sum_{\{s,k\, \| \, 
\overline{{\cal B}}\}} \int \!\!  D_{\,\overline{{\cal B}}}[\, \overline{\psi}, \psi ]  \; 
\prod_{x} \frac{ \Big( 2m \,  \overline{\psi}_x \psi_x \Big)^{s_x}\!\!}{s_x !} \;
\label{Zcomplement}
\\
&& \hspace{10mm} \times \;
\prod_{x,\nu} 
\frac{(3-k_{x,\nu})!}{6 \,  k_{x,\nu}!} \big(\,\overline{\psi}_x  \psi_x \;\;
\overline{\psi}_{x  +  \hat{\nu}} \psi_{x + \hat{\nu}}\big)^{k_{x,\nu}} \! .
\nonumber
\end{eqnarray}
$Z_{\,\overline{{\cal B}}}$ is a sum over all possibilities to fill the complementary domain $\overline{{\cal B}}$ domain with 
quark and diquark monomers and quark and diquark dimers. Note that all corresponding insertions of Grassmann monomials
contribute the Grassmann variables already in the canonical order such that all contributions to $Z_{\overline{{\cal B}}}$ 
come with real and positive factors. The corresponding weights can easily be determined from (\ref{Zcomplement}): A quark monomer 
$\overline{\psi}_{x,a} \, \psi_{x,a}$ comes with a factor of $2m$, while a diquark monomer   
$\overline{\psi}_{x,a} \, \psi_{x,a} \; \overline{\psi}_{x,b} \, \psi_{x,b}$, $a \neq b$ comes with $(2m)^2$ (the factor 1/2! for $s_x = 2$ 
in (\ref{Zcomplement}) is cancelled
because the diquark monomer is a mixed term in a square and thus comes with a factor of 2). A quark dimer 
$\overline{\psi}_{x,a} \, \psi_{x,a} \; \overline{\psi}_{x+\hat{\nu},c} \, \psi_{x+\hat{\nu},c}$ has a weight of $1/3$ and the same weight 
also results for a diquark dimer $\overline{\psi}_{x,a} \, \psi_{x,a} \; \overline{\psi}_{x,b} \, \psi_{x,b} \;  
\overline{\psi}_{x+\hat{\nu},c} \, \psi_{x+\hat{\nu},c} \; \overline{\psi}_{x+\hat{\nu},d} \, \psi_{x+\hat{\nu},d}$ 
(here we have a product of two mixed terms 
which gives a factor of 4 that combines with the factor 1/12 that comes from (\ref{Zcomplement}) when setting $k_{x,\nu} = 2$, to the factor of 1/3
we quote).

With these weights we can write the contribution $Z_{\,\overline{{\cal B}}}$ for the complementary domain in the following form:
\begin{equation}
Z_{\,\overline{{\cal B}}} \; =  \! \sum_{\{m,d\, \| \, \overline{{\cal B}}\}}  \!
(2m)^{{\cal N}(m_1) + {\cal N}(m_2)} \left( \frac{1}{3} \right)^{{\cal N}(d_1) + {\cal N}(d_2)},
\label{ZBfinal}
\end{equation}
where the sum runs over all possible configurations of monomer and dimer configurations in the complementary domain 
$\overline{{\cal B}}$ according to the rules illustrated in Fig.~\ref{complementary}. By ${\cal N}(m_1)$ and ${\cal N}(m_2)$ we denote
the total number of quark and diquark monomers and by ${\cal N}(d_1)$ and ${\cal N}(d_2)$ the total number of quark and diquark dimers.

\section{Fermion determinants for baryon bag contributions}

We finally show that the contribution $Z_{{\cal B}_i}$  of a baryon bag ${\cal B}_i$ can be written 
as the determinant\footnote{We stress at this point that of course the fermionic Gaussian integral can always be written as 
a determinant, but the factorized form as a product over bag determinants in Eq.~(43) is possible only in the strong coupling limit, which suppresses off-diagonal terms that connect different baryon bags.}
of the Dirac operator $D^{(i)}$ in the fermion bag as defined in (\ref{bagoperator}). The first step is to  reorganize
the Grassmann measure (\ref{measure_bag}) inside the baryon bag ${\cal B}_i$:
\begin{eqnarray}
D_{{\cal B}_i}[\, \overline{\psi}, \psi ]  & = & 
\prod_{x \in {\cal B}_i} \prod_a  d\psi_{x,a} \, d\overline{\psi}_{x,a} 
\nonumber \\
& = & 
\prod_{x \in {\cal B}_i}  d\psi_{x,3} \, d\psi_{x,2} \, d\psi_{x,1} \, 
d\overline{\psi}_{x,1} \, d\overline{\psi}_{x,2} \, d\overline{\psi}_{x,3} 
\nonumber \\
& = & 
\prod_{x \in {\cal B}_i}  dB_x \, d\overline{B}_x \; ,
\end{eqnarray}
where we have defined the baryon measures
\begin{eqnarray}
dB_x &\;  \equiv \; & d\psi_{x,3} \, d\psi_{x,2} \, d\psi_{x,1} \; , 
 \\
d\overline{B}_x & \; \equiv \; & d\overline{\psi}_{x,1} \, d\overline{\psi}_{x,2} \, d\overline{\psi}_{x,3} \; .
\end{eqnarray}
Together with the baryon fields $B_x$ and $\overline{B}_x$ defined in (\ref{baryondefs}) they obey the following anti-commutation
relations
\begin{eqnarray}
\hspace*{-8mm} && \{  B_x , B_y \} =  0 , \, \{  \overline{B}_x , \overline{B}_y \} = 0 , \, \{  B_x , \overline{B}_y \} = 0 ,
\\
\hspace*{-8mm} && \{  dB_x , dB_y \} = 0 , \, \{  d\overline{B}_x , d\overline{B}_y \} = 0 , \, \{  dB_x , d\overline{B}_y \} = 0,
\\
\hspace*{-8mm} && \{  dB_x , B_y \} =  0 , \, \{  dB_x , \overline{B}_y \} = 0 ,
\\
\hspace*{-8mm} && \{  d\overline{B}_x , B_y \} = 0 , \, \{  d\overline{B}_x , \overline{B}_y \} = 0 .
\end{eqnarray}
Furthermore the baryon measures and the baryon fields obey the Grassmann integration rules
\begin{eqnarray}
\hspace*{-8mm} && \int dB_x \, 1 \, = \, 0 \, , \; \int dB_x \, B_x \, = \, 1 \, ,
\\
\hspace*{-8mm} && \int d\overline{B}_x  \, 1 \, = \, 0 \, , \; \int d\overline{B}_x  \, \overline{B}_x \, = \, 1 \, ,
\end{eqnarray}
which they inherit from the underlying properties of the Grassmann variables $\psi_{a,x}$, $\overline{\psi}_{a,x}$, 
$d\psi_{a,x}$, $d\overline{\psi}_{a,x}$ they are made of. Thus the baryon fields $B_x$ and $\overline{B}_x$, together
with the measures $dB_x$ and $d\overline{B}_x$ form a representation of a Grassmann algebra and the corresponding measures.

As a consequence we can identify the contribution $Z_{{\cal B}_i}$ of the baryon bag ${\cal B}_i$ as a 
Gaussian Grassmann integral and thus a determinant,
\begin{equation}
Z_{{\cal B}_i}  \, =  \int \!\! \prod_{x \in {\cal B}_i} \!\! dB_x \, \overline{B}_x \; \exp\!\left( \! \sum_{x,y} \overline{B}_x \, D^{(i)}_{\;x,y} \, B_y \! \right)  
\, = \, \det D^{(i)} ,
\end{equation}
where $D^{(i)}$ is the baryon bag Dirac operator defined in (\ref{bagoperator}) for the baryons in the bag ${\cal B}_i$.

For vanishing chemical potential, i.e., $\mu = 0 = \mu_B$, the properties of the Dirac operator under chiral transformations 
guarantee the positivity of the determinant: We write the baryon bag Dirac operator $D^{(i)}$ in the form
\begin{eqnarray}
D^{(i)} \; = \; 2M \,  \mathbbm{1}^{(i)} \; + \; A^{(i)} \; ,
\nonumber
\end{eqnarray}
where $\mathbbm{1}^{(i)}$ is the unit operator on the set of sites inside the bag ${\cal B}_i$, and 
$A^{(i)}$ denotes the hopping terms of $D^{(i)}$ at $\mu_B = 0$ according to the definition (\ref{bagoperator}).
The matrix $A^{(i)}$ is anti-hermitian, such that it has purely imaginary eigenvalues $i \lambda$ with $\lambda \in \mathbb{R}$.
Furthermore $A^{(i)}$ anti-commutes with
\begin{equation}
\Gamma_{5 \; \; x,y}^{(i)} \; = \; \theta^{(i)}_x \; (-1)^{x_1+x_2+x_3+x_4} \; \delta_{x,y} \; ,
\end{equation}
which is the staggered representation of $\gamma_5$ restricted to the bag ${\cal B}_i$. Thus at $\mu_B = 0$ the eigenvalues of 
$D^{(i)}$ come in complex conjugate pairs $2M \pm i \lambda$ such that their product $4 M^2 + \lambda^2$ is non-negative 
(strictly positive for $M \neq 0$) and thus $\det D^{(i)}$ is non-negative (strictly positive). 

We stress however, that for $\mu_B \neq 0$ this argument does not hold, and $\det D^{(i)}$ can be negative. A bag may, e.g., contain a
single baryon loop that winds around compact time with a negative sign. For $\mu_B = 0$ this negative contribution is compensated by the 
bayron dimers and monomers, but for $\mu_B$ this loop has an additional factor of $\cosh(\mu_B \, N_t)$, such that for sufficiently large
$\mu_B \, N_t$ the contribution $Z_{{\cal B}_i} = \det D^{(i)}$ will be negative. Thus also in the baryon bag representation strong
coupling QCD has the sign problem of free fermions inside the bags.

We conclude this section with summarizing the final form of the baryon bag representation for the partition sum of strong
coupling QCD:
\begin{eqnarray}
Z \; =  \; \sum_{\{{\cal B}\}}  \prod_i \det \, D^{(i)} \times \; Z_{\overline{{\cal B}}} \; .
\label{zstrongfinal}
\end{eqnarray}
The partition function $Z$ is the sum $\sum_{\{{\cal B}\}}$ over all configurations of baryon bags. For each baryon bag 
${\cal B}_i$ we pick up a weight factor $\det \, D^{(i)}$ given by the fermion determinant of the baryon bag Dirac operator (\ref{bagoperator}).
These determinants are real and positive for vanishing chemical potential, but may become negative for sufficiently large $\mu_B$. 
In the complementary domain $\overline{{\cal B}}$ the dynamics is described by quark and diquark monomer and dimer terms. The
corresponding factor  $Z_{\overline{{\cal B}}}$ is given in (\ref{ZBfinal}) and is always real and positive.

\section{Concluding remarks}

In this paper we have analyzed the baryon contributions in lattice QCD with one flavor of staggered fermions. Using Taylor expansion of the 
Boltzmann factors for the individual site- and link terms of the fermion action we could separate the corresponding baryon contributions. 
Re-exponentiating these terms we were able to organize their contributions in the form of a separate Boltzmann factor for the baryons. The baryon 
fields are monomials of three quark fields and no longer couple to the gauge field. The action for the baryons turns out to be the free staggered 
action with anti-periodic temporal boundary conditions and for the case of vanishing quark mass the action is chiral. 

We stress that the separation of the baryon terms with a Boltzmann factor for free fermions holds for arbitrary values of the gauge coupling 
and does not depend on the strong coupling limit. 
We expect that the form with factorized baryon contributions will be very useful for simplifying worldline/worldsheet 
representations of lattice QCD. We keep this for future work and in this paper continued with considering the strong coupling limit. 

In the strong coupling limit the Grassmann measure could be factorized into baryon bags where the Grassmann integral is separated 
with terms from the baryon Boltzmann factor and a complementary domain where the relevant degrees of freedom are monomers and 
dimers for quarks and diquarks. Inside the baryon bags the Grassmann measure was rewritten as a measure for the baryons and we found 
that the baryon measure and the baryon fields together obey a Grassmann algebra. Inside each baryon bag the Gaussian integral for the baryons 
gives rise to a fermion determinant for the Dirac operator restricted to the respective bag. In the complementary domain the quark and diquark 
contributions always have real and positive weights and for the baryon bags the chiral transformation properties of the Dirac operator ensure
that for vanishing chemical potential $\mu$ the baryon bag determinants are real and positive. For nonzero $\mu$ the baryon bag 
representation has the sign problem of free fermions which becomes manifest once the fermion bags reach the size of the inverse temperature. 

The result (\ref{zstrongfinal}) for the strong coupling partition sum has the form of a fermion bag representation \cite{FBrev1,FBrev2}, but 
conceptually the underlying structure is more complex than previous fermion bag representations:  Here the fermion terms inside the baryon 
bags describe the joint propagation of three quarks and thus do not emerge from fundamental fermion fields in the action. In the same way 
the interaction terms in the complementary domain are not directly generated from expanding the Boltzmann factor for a genuine interaction term 
in the action, but instead emerge from an interplay of quark and diquark terms. 

Within the framework of strong coupling QCD, the identification of baryon bags gives rise to a new type of resummation of configurations 
in the strong coupling representation \cite{RossiWolff,KarschMutter}. We stress, however, that the resummation of baryon contributions in a 
baryon bag determinant is different from other forms of resummation where single baryon loops are combined with chains of dimers in
individual update steps \cite{RossiWolff,KarschMutter}.

We conclude with stressing that the partition function is a sum over all bag configurations, and the configurations of bags come with 
different weights that depend on the parameters, e.g., mass, temperature and chemical potential. Thus in a Monte Carlo simulation the 
average size and the total number of bags per volume will depend on these parameters. Consequently the baryon bag representation 
allows the system to dynamically choose the combinations of degrees of freedom that are relevant for a given set of couplings, either in terms of baryons, 
which will give rise to a large fraction of sites belonging to baryon bags, or in terms of quarks and diquarks which corresponds to a large 
fraction of sites in the complementary domain. It will be interesting to explore this dynamical selection of the relevant degrees of freedom
in a numerical simulation of the baryon bag representation.

\begin{acknowledgments}
We thank Bastian Brandt and Carlotta Marchis for interesting discussions.
This work is supported by the Austrian Science Fund FWF, grant I 2886-N27, as well as 
DFG TR55, ''$\!$Hadron Properties from Lattice QCD''.	
\end{acknowledgments}



\begin{thebibliography}{12}
  
\bibitem{rev1}
  S.~Chandrasekharan,
  PoS LATTICE {\bf 2008} (2008) 003
  [arXiv:0810.2419].

\bibitem{rev2}
  P.~de Forcrand,
  PoS LAT {\bf 2009} (2009) 010
  [arXiv:1005.0539].

\bibitem{rev3}
 U.~Wolff,
  PoS LATTICE {\bf 2010} (2010) 020
  [arXiv:1009.0657].
  
\bibitem{rev4}
C.~Gattringer,
PoS LATTICE {\bf 2013} (2013) 002
[arXiv:1401.7788].

\bibitem{RossiWolff}
  P.~Rossi and U.~Wolff,
  Nucl.\ Phys.\ B {\bf 248} (1984) 105.

\bibitem{KarschMutter}
  F.~Karsch, K.H.~M{\"u}tter,
  Nucl.\ Phys.\ B {\bf 313} (1989) 541.

\bibitem{scQCD1}
D.~H.~Adams and S.~Chandrasekharan,
  Nucl.\ Phys.\ B {\bf 662}, 220 (2003)
  [hep-lat/0303003].

\bibitem{scQCD2}
  S.~Chandrasekharan and F.~J.~Jiang,
  Phys.\ Rev.\ D {\bf 68}, 091501 (2003)
  [hep-lat/0309025].

\bibitem{scQCD3} 
S.~Chandrasekharan and C.~G.~Strouthos,
  Phys.\ Rev.\ D {\bf 69} (2004) 091502
  [hep-lat/0401002].

\bibitem{scQCD4} 
  P.~de Forcrand and M.~Fromm,
  Phys.\ Rev.\ Lett.\  {\bf 104}, 112005 (2010)
  [arXiv:0907.1915 [hep-lat]].
  
\bibitem{scQCD5} 
  W.~Unger and P.~de Forcrand,
  J.\ Phys.\ G {\bf 38}, 124190 (2011)
  [arXiv:1107.1553 [hep-lat]].
  
\bibitem{scQCD6} 
  P.~de Forcrand, S.~Kim and W.~Unger,
  JHEP {\bf 1302}, 051 (2013)
  [arXiv:1208.2148 [hep-lat]].
  
 \bibitem{scQCD7}  
 P.~de Forcrand, J.~Langelage, O.~Philipsen and W.~Unger,
  Phys.\ Rev.\ Lett.\  {\bf 113}, no. 15, 152002 (2014)
  [arXiv:1406.4397 [hep-lat]].
  
\bibitem{scQCD8}
  W.~Unger,
  PoS LATTICE {\bf 2014} (2014) 192
  [arXiv:1411.4493 [hep-lat]].

\bibitem{scQCD9}
J.~Kim and W.~Unger,
  PoS LATTICE {\bf 2016}, 035 (2016)
  [arXiv:1611.09120 [hep-lat]].

\bibitem{scQCD10} 
  P.~de Forcrand, P.~Romatschke, W.~Unger and H.~Vairinhos,
  PoS LATTICE {\bf 2016}, 086 (2017)
  [arXiv:1701.08324 [hep-lat]].
  
\bibitem{scQCD11}
  P.~de Forcrand, W.~Unger and H.~Vairinhos,
  arXiv:1710.00611 [hep-lat].
 
\bibitem{GM2} 
  C.~Marchis and C.~Gattringer,
  PoS LATTICE {\bf 2016}, 034 (2016)
  [arXiv:1611.01022 [hep-lat]].

\bibitem{Gagliardi} 
  G.~Gagliardi, J.~Kim and W.~Unger,
  arXiv:1710.07564 [hep-lat].
   
\bibitem{GM3} 
  C.~Gattringer, D.~G{\"o}schl and C.~Marchis,
  arXiv:1710.08745 [hep-lat].

\bibitem{Borisenko}  
  O.~Borisenko, V.~Chelnokov and S.~Voloshyn,
  arXiv:1712.03064 [hep-lat].
 
\bibitem{GM4}
 C.~Marchis and C.~Gattringer,
  Phys.\ Rev.\ D {\bf 97} (2018) 034508
  [arXiv:1712.07546 [hep-lat]].
  
\bibitem{FBrev1}
S.~Chandrasekharan,
  Phys.\ Rev.\ D {\bf 82}, 025007 (2010)
  [arXiv:0910.5736 [hep-lat]].

\bibitem{FBrev2}
S.~Chandrasekharan,
  Eur.\ Phys.\ J.\ A {\bf 49} (2013) 90
  [arXiv:1304.4900 [hep-lat]].

\bibitem{FBres1}
S.~Chandrasekharan and A.~Li,
  JHEP {\bf 1101}, 018 (2011)
  [arXiv:1008.5146 [hep-lat]].

\bibitem{FBres2}  
S.~Chandrasekharan and A.~Li,
  Phys.\ Rev.\ Lett.\  {\bf 108}, 140404 (2012)
  [arXiv:1111.7204 [hep-lat]].
   
\bibitem{FBres3} 
S.~Chandrasekharan and A.~Li,
  Phys.\ Rev.\ D {\bf 85}, 091502 (2012)
  [arXiv:1202.6572 [hep-lat]].
   
\bibitem{FBres4}
S.~Chandrasekharan,
  Phys.\ Rev.\ D {\bf 86}, 021701 (2012)
  [arXiv:1205.0084 [hep-lat]].

\bibitem{FBres5}
S.~Chandrasekharan and A.~Li,
  Phys.\ Rev.\ D {\bf 88}, 021701 (2013)
  [arXiv:1304.7761 [hep-lat]].

\bibitem{FBres6}
 E.~F.~Huffman and S.~Chandrasekharan,
  Phys.\ Rev.\ B {\bf 89}, 111101 (2014)
  [arXiv:1311.0034 [cond-mat.str-el]].
 
 \bibitem{FBres7}
 V.~Ayyar and S.~Chandrasekharan,
  Phys.\ Rev.\ D {\bf 91}, 065035 (2015)
  [arXiv:1410.6474 [hep-lat]].
 
\bibitem{FBres8}
  V.~Ayyar and S.~Chandrasekharan,
  Phys.\ Rev.\ D {\bf 93},  081701 (2016)
  [arXiv:1511.09071 [hep-lat]].

\bibitem{FBres9}
 V.~Ayyar and S.~Chandrasekharan,
  JHEP {\bf 1610}, 058 (2016)
  [arXiv:1606.06312 [hep-lat]].
 
\bibitem{FBres10}  
E.~Huffman and S.~Chandrasekharan,
  Phys.\ Rev.\ D {\bf 96}, 114502 (2017)
  [arXiv:1709.03578 [hep-lat]].

\bibitem{FBres11}
  V.~Ayyar, S.~Chandrasekharan and J.~Rantaharju,
  arXiv:1711.07898 [hep-lat].

\bibitem{creutz}
M.~Creutz,
J.\ Math.\ Phys.\  {\bf 19}, 2043 (1978).
    
\end{thebibliography}

\end{document}